\documentclass[twocolumn]{revtex4}

\usepackage{graphicx}
\usepackage{amsmath}
\usepackage{color}
\usepackage{ulem}

\begin{document}
\title{Electric dipole moment searches: reexamination of frequency shifts for particles in traps}
\author{Guillaume Pignol}
\email{guillaume.pignol@lpsc.in2p3.fr} 
\affiliation{LPSC, Universit\'{e} Joseph Fourier - CNRS/IN2P3 - INPG, 53 rue des Martyrs, F-38026 Grenoble, France}
\author{St\'ephanie Roccia}
\email{roccia@csnsm.in2p3.fr} 
\affiliation{CSNSM, Universit\'e Paris Sud - CNRS/IN2P3, b\^at. 104 et 108, F-91405 Orsay-campus, France}

\date{\today}

\begin{abstract}
In experiments searching for a non-zero electric dipole moment of trapped particles, 
frequency shifts correlated with an applied electric field can be interpreted as a false signal. 
One such effect, referred to as the geometric phase effect, is known to occur in a magnetic field that is non-perfectly homogeneous. 
The increase in sensitivity of experiments demands improved theoretical description of this effect. 
In the case of fast particles, like atoms at room temperature and low pressure, 
the validity of established theories was limited to a cylindrical confinement cell in a uniform gradient with cylindrical symmetry. 
We develop a more general theory valid for an arbitrary shape of the magnetic field as well as for arbitrary geometry of the confinement cell. 
Our improved theory is especially relevant for experiments measuring the neutron electric dipole moment with an atomic co-magnetometer. 
In this context, we have reproduced and extended earlier numerical studies of the geometric phase effect induced by localized magnetic impurities. 
\end{abstract}

\maketitle

\section{Introduction}

A non-zero electric dipole moment (EDM) for a spin 1/2 particle would violate both the parity (P) and time reversal (T) symmetry, it would also violate the CP symmetry according to the CPT connection. 
Since CP violation beyond the standard model (SM) is needed to explain the generation of the baryon asymmetry in the early universe, electric dipole moments are important inputs for cosmological models and theories beyond the SM \cite{Khriplovich}. 
In particular, the neutron EDM has been shown to strongly constrain many such theories \cite{Dubbers, Pospelov}. 
So far, the best measurement of the neutron EDM \cite{Baker}, extracted from the precession frequency of trapped ultracold neutrons, is compatible with zero. 

When designing modern experiments to search for a non zero electric dipole moment of the neutron, the control of the magnetic field fluctuations is of great significance. 
A successful strategy is the use of an atomic comagnetometer, where polarized atoms fill the neutron confinement volume. 
In this volume, a combination of a weak magnetic field $B_0$ and strong electric field $E$ is applied, either in a parallel or an antiparallel configuration. 
The precession frequency of the comagnetometer measures the magnetic field via the first order relation $\omega = \gamma B_0$, where $\gamma$ is the gyromagnetic ratio of the atom. 
The comagnetometer measurement is then used to correct the neutron spin-precession frequency for the magnetic field fluctuations as in the performed experiment \cite{Baker} using a mercury comagnetometer \cite{mercury}. 
In turn, the comagnetometer could induce false effects in the extraction of the neutron EDM. 
The most dangerous effects are frequency shifts $\delta \omega(E)$ of the comagnetometer correlated with the direction and value of the electric field. 
In this case, a false EDM for the comagnetometer is generated by the $E$-odd component of the frequency shift
\begin{equation}
d_{\rm False} = \frac{\hbar}{4 E} \left( \delta \omega(E) - \delta \omega(-E) \right)
\end{equation}
that transmits into a false neutron EDM $\Delta d_n = \gamma_n/\gamma \ d_{\rm False}$, where $\gamma_n$ is the neutron gyromagnetic ratio. 
In particular, it has been recognized both theoretically and experimentally that such effects arise from geometric phase \cite{Commins,Pendlebury}: 
when a particle moves with a velocity ${\bf v}$ with respect to an electric field ${\bf E}$, it sees effectively a motional magnetic field ${\bf E} \times {\bf v} / c^2$. 
As a result, each spin-polarized particle  
is submitted to a fluctuating field (that we will refer to as $\boldsymbol \omega / \gamma$) resulting from a combination of the magnetic field inhomogeneities $(B_x, B_y, B_z)$ and the motional field. 
Within this article we assume that the main magnetic field $B_0 = \omega_0 / \gamma$ and the electric field are aligned with the $z$ axis. 
The fluctuating transverse components of $\boldsymbol \omega$
\begin{eqnarray}
\label{omega}
\omega_x & = & \gamma B_x - \gamma \frac{E}{c^2} v_y \nonumber \\
\omega_y & = & \gamma B_y + \gamma \frac{E}{c^2} v_x
\end{eqnarray}
are known to induce a \textit{geometric phase} shift in the spin-precession frequency 
(in the case of a rotating transverse field, the frequency shift is also know as the Ramsey-Bloch-Siegert shift). 

The particular case of a uniform magnetic field gradient with cylindrical symmetry
\begin{equation}
\label{cylindricalGradient}
B_x = - \frac{x}{2} \frac{\partial B_z}{\partial z}, \quad B_y = - \frac{y}{2} \frac{\partial B_z}{\partial z}
\end{equation}
has been considered in \cite{Pendlebury}. 
In the low field regime, the resulting $E$-odd frequency shift has been derived using an elementary approach: 
\begin{equation}
\delta \omega_{BE} = \frac{\gamma^2 E}{16 c^2} D^2 \, \frac{\partial B_z}{\partial z}.
\end{equation}
The calculation is done assuming a cylindrical trap of diameter $D$ with the axis of the cylinder aligned with the $z$ axis. 
The approach used in \cite{Pendlebury} is based on solving explicitly the Bloch equations for specific particle trajectories 
and could not be generalized to the case of arbitrary shape of the transverse magnetic field. 
Then a more powerful theoretical treatment using the general theory of relaxation developed in \cite{LamoreauxGolub,BarabanovGolubLamoreaux} gave new insight to the problem, 
however still focusing on the case of a cylindrical uniform field gradient. 
In this article we review and extend the general approach \cite{LamoreauxGolub,BarabanovGolubLamoreaux}, 
with the goal to apply it to the case of the mercury comagnetometer \cite{mercury} used in the EDM experiment  \cite{EDM}. 
The magnetometer runs in the ballistic regime (gas collisions much less frequent than wall collisions) and in the nonadiabatic low field regime (the wall collision rate is much faster than the Larmor frequency) for which no theory was available beyond the case of a cylindrical cell in uniform magnetic field gradients with a cylindrical symetry. 
As a main result of the present work, we establish the expression for the $E$-odd frequency shift valid for arbitrary shape of the magnetic field and for arbitrary shape of the trap: 
\begin{equation}
\label{mainresultBE}
\delta \omega_{BE} = - \frac{\gamma^2 E}{c^2} \langle x B_x + y B_y \rangle
\end{equation}
where the brackets refer to the volume average over the trap.
The case of an arbitrary field shape has been considered in \cite{Clayton} for the case of a rectangular cell, 
with focus on the case where the particles are diffusing by gas collisions (high density limit). 

We first recall the general approach for the frequency shift  
and establish the terms proportional to $E$, $E^2$, $B^2$ respectively, valid for any shape of the magnetic inhomogeneity and arbitrary shape of the trap. 
Then we discuss implications for neutron EDM experiments using an atomic comagnetometer in the nonadiabatic regime (such as the mercury comagnetometer). 
Whenever the geometry of the trap has to be specified for numerical studies, we will consider that of the experiment \cite{EDM} (the \textit{OILL trap}), 
a cylindrical trap whose axis is aligned with the main magnetic field, with a diameter $D=47$~cm and a height $H = 12$~cm.

\section{General results for frequency shifts}

The frequency shift induced by a fluctuating transverse field is given by the Lamoreaux-Golub expression \cite{LamoreauxGolub}: 
\small
\begin{eqnarray}
\label{generalFormula}
\delta \omega = \frac{1}{2} \int_0^\infty d \tau 
\cos(\omega_0 \tau) \ \langle \omega_x(0) \omega_y(\tau) - \omega_y(0) \omega_x(\tau) \rangle \\
\nonumber
 + \frac{1}{2} \int_0^\infty d \tau 
\sin(\omega_0 \tau) \ \langle \omega_x(0) \omega_x(\tau) + \omega_y(0) \omega_y(\tau) \rangle.
\end{eqnarray}
\normalsize
where the sign of the second term has been corrected (see appendix \ref{derivation} for a derivation of this expression). 
Here the brackets refer to the ensemble average of the quantity over all particles in the trap. 
In our case where an electric field is present, the transverse field is given by (\ref{omega}) and the general formula (\ref{generalFormula}) 
should be expanded in powers of the electric field: 
\begin{equation}
\delta \omega = \delta \omega_{B^2} + \delta \omega_{E^2}  + \delta \omega_{BE}.
\end{equation}
The $E$-independent frequency shift reads: 
\small
\begin{equation}
\delta \omega_{B^2} = \frac{\gamma^2}{2} \int_0^\infty d \tau 
\sin(\omega_0 \tau) \ \langle  B_x(0) B_x(\tau) + B_y(0) B_y(\tau) \rangle.
\end{equation}
\normalsize
It does not induce a direct false electric dipole moment since it remains unchanged when the electric field is reversed. 
However, it has to be taken into account in the EDM analysis procedure.
The quadratic frequency shift reads: 
\small
\begin{equation}
\delta \omega_{E^2} = \frac{\gamma^2 E^2}{2 c^4} \int_0^\infty d \tau 
\sin(\omega_0 \tau) \ \langle v_x(0) v_x(\tau) + v_y(0) v_y(\tau) \rangle.
\end{equation}
\normalsize
It can induce a false electric dipole moment if the magnitude of the E field changes when it is reversed \cite{Lamoreaux}. 

Finally, the most important frequency shift is the $E$-odd term involving a linear dependence in the electric field: 
\small
\begin{equation}
\label{generalBE}
\delta \omega_{BE} = \frac{\gamma^2 E}{c^2} \int_0^\infty d \tau \cos(\omega_0 \tau) \ \langle B_x(0) v_x(\tau) + B_y(0) v_y(\tau) \rangle.
\end{equation}
\normalsize
It generates a false electric dipole moment for the comagnetometer: 
\begin{equation}
\label{dFalse}
d_{\rm False} = \frac{\hbar}{4 E} \left( \delta \omega(E) - \delta \omega(-E) \right) = \frac{\hbar}{2 E} \delta \omega_{BE}(E)
\end{equation}


As we have just shown, the frequency shifts $\delta \omega_{B^2}, \delta \omega_{E^2}, \delta \omega_{BE}$ involve Fourier transform (evaluated at the Larmor frequency) 
of correlation functions involving field and velocity components. 
In the abiabatic regime the Fourier transforms can be expanded in perturbation series using integration by part: 
\small
\begin{eqnarray}
\label{ipp}
& \int_0^\infty d \tau \cos(\omega_0 \tau) \ \langle B_x(0) v_x(\tau) \rangle = \\ 
\nonumber
& \left[ \cos(\omega_0 \tau) \langle B_x(0) x(\tau) \rangle \right]_0^\infty + \omega_0 \int_0^\infty d \tau \sin(\omega_0 \tau) \ \langle B_x(0) x(\tau) \rangle
\end{eqnarray}
\normalsize
where the second term vanishes in the nonadiabatic limit ($\omega_0 \tau_c \ll 1$). 
Using the previous expansion (\ref{ipp}) in (\ref{generalBE}) we arrive at the expression (\ref{mainresultBE}), which translates into a false EDM: 
\begin{equation}
\label{dFalseHg}
d_{\rm False} = - \frac{\hbar \gamma^2}{2 c^2} \langle x B_x + y B_y \rangle
\end{equation}
valid for arbitrary field inhomogeneities in the nonadiabatic regime. 

In the presence of a cylindrical uniform gradient (\ref{cylindricalGradient}) in a cylindrical trap, one has to calculate the volume average
\begin{equation}
\langle x^2+y^2 \rangle = \frac{4}{\pi D^2} \int_0^{D/2} r^2 \ 2 \pi r dr = \frac{D^2}{8}
\end{equation}
to show that the general result (\ref{dFalseHg}) reduces to: 
\begin{equation}
\label{14}
d_{\rm False} =  \frac{\hbar \gamma^2}{32 c^2} \ D^2 \frac{\partial B_z}{\partial z}
\end{equation}
which coincides with eq. (37) of \cite{Pendlebury}. 
Let us point out that, although the full correlation function depends on the details of the random particle motion (such as the degree of specularity of reflection at wall collisions), 
this dependence disappears in the nonadiabatic limit, where only  the volume average of field components is involved. 
The same conclusion was arrived at in \cite{Pendlebury} in the case of a cylindrical uniform gradient, using numerical simulations. 

%

\section{Implications of the linear frequency shift}

Let us now work out the implications of the linear frequency shift of the comagnetometer when extracting the neutron electric dipole moment. 

\subsection{Large scale magnetic inhomogeneities}

First we deal with large scale magnetic inhomogeneities arising for, for example, non-perfect symmetry of the coil generating $B_0$ 
or non-perfect shielding of the external ambient field. 
It is useful to parametrize the field components by polynomials of $x,y,z$ up to a certain order. 
These polynomials however cannot be arbitrary since the Maxwell equations must hold. 
A convenient way to find a minimal polynomial expansion is to consider a scalar magnetic potential $\phi$ for the field: 
\begin{equation}
{\bf B}(x,y,z) = {\bf grad} \ \phi (x,y,z).
\end{equation}
In order for ${\bf B}$ to satisfy Maxwell's equations, it is necessary and sufficient for $\phi$ to satisfy the Laplace equation $\Delta \phi = 0$. 
Then, the general first order inhomogeneities are described by the following second order harmonic polynomials: 
\small
\begin{eqnarray}
\nonumber
\phi(x,y,z) =	   B_0 z + \frac{G_x}{2} x^2 + \frac{G_y}{2} y^2 - \frac{1}{2}(G_x + G_y) z^2 \\ 
		 + Q_x y z + Q_y x z + Q_z x y.
\end{eqnarray}
\normalsize
The inhomogeneous part has in total 5 degrees of freedom. 
It corresponds to the transverse fields: 
\begin{eqnarray}
\label{generalGradient}
\nonumber
B_x & = & G_x x + Q_y z + Q_z y \\
B_y & = & G_y y + Q_x z + Q_z x
\end{eqnarray}
This \textit{general uniform gradient} can be used to fit a magnetic map in the experiment \cite{EDM} 
and is much more general than the sole cylindrical uniform gradient (\ref{cylindricalGradient}) considered in earlier studies \cite{Pendlebury}. 
Then, from eq. (\ref{dFalseHg}) we arrive at the expression for the false EDM of the mercury atoms: 
\begin{equation}
d_{\rm False} = - \frac{\hbar \gamma^2}{32 c^2} \ D^2 \left( G_x + G_y \right).
\end{equation}
Given that in this case $G_x + G_y = - \langle \frac{\partial B_z}{\partial z} \rangle$, 
the previous expression is in fact no different from that of a cylindrical uniform gradient. 


It suggests the following strategy of data analysis, which in fact was used in \cite{Baker}: 
a correction of the measured EDM is applied to subtract the geometric phase shift of the mercury atoms, according to: 
\begin{equation}
\label{Corr}
d_{\rm corr} = d_{\rm meas} - \frac{\hbar \gamma_n \gamma_{\rm Hg}}{32 c^2} \ D^2 \langle \frac{\partial B_z}{\partial z} \rangle, 
\end{equation}
the volume average magnetic field gradient $\langle \frac{\partial B_z}{\partial z} \rangle$ being evaluated independently. 
This strategy is reinforced by the new result presented here, which shows that the correction is valid for the general uniform gradient (\ref{generalGradient})
and not only for the case of cylindrical uniform gradient (\ref{cylindricalGradient}).

\subsection{Small scale magnetic inhomogeneities}

We have shown that the correction (\ref{Corr}) is efficient at removing the false geometric phase effect induced by large scale magnetic inhomogeneities, 
but as already noticed in \cite{Baker,HarrisPendlebury}, it could become inefficient in the presence of a localized magnetic inhomogeneity where the resulting field gradients are not uniform over the trap volume.

\begin{figure}
\begin{center}
\includegraphics[angle=90,width=0.98\linewidth]{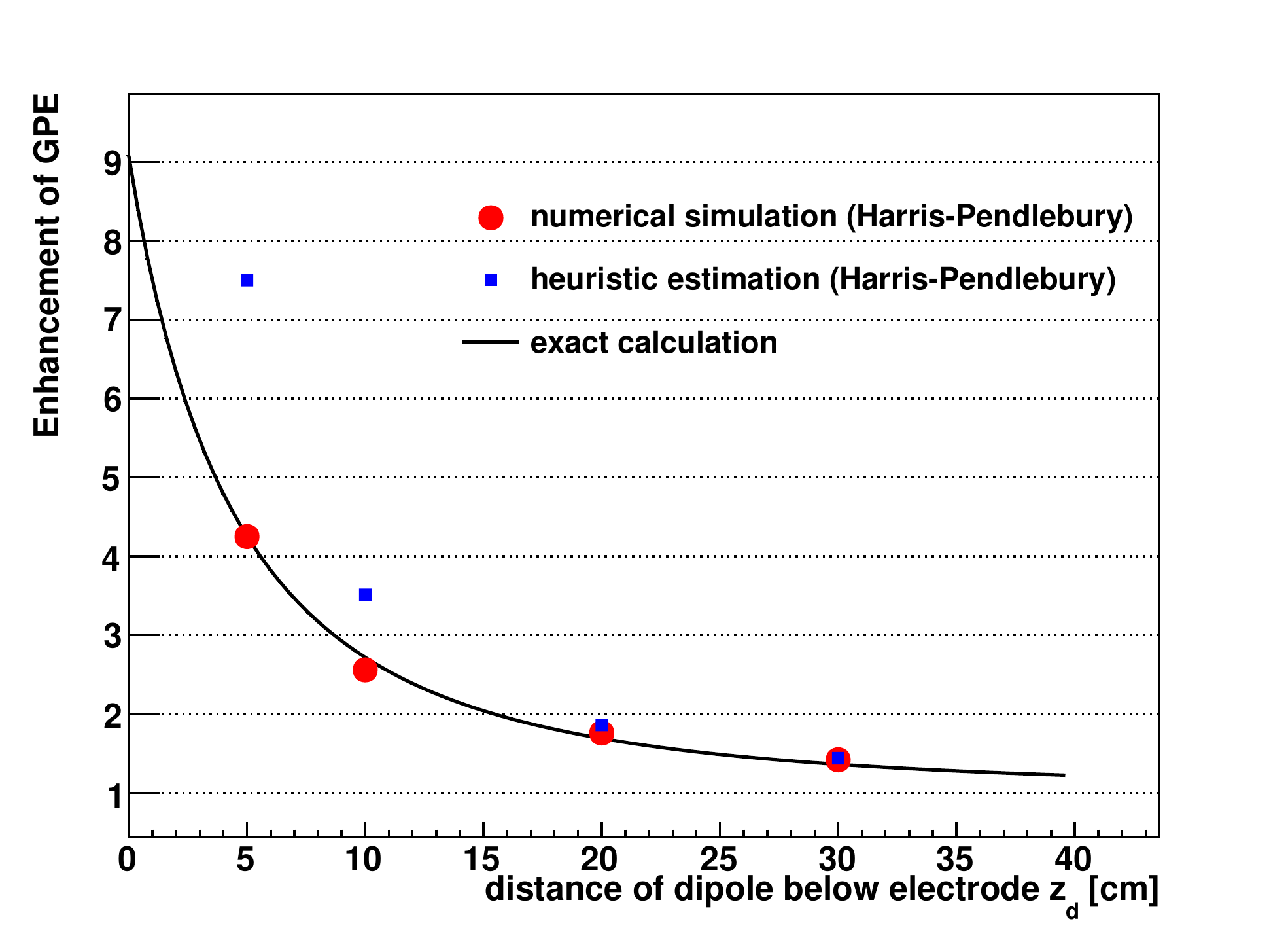}
\caption{
Enhancement of the geometric phase effect calculated for the OILL trap for a magnetic dipole on axis 
as a function of the distance from the bottom plate of the trap. 
The analytical result (\ref{AnalyticDipole}) is compared to the heuristic and numerical results \cite{HarrisPendlebury}. 
\label{compHarrisPendlebury}
}
\end{center}
\end{figure}

The correction (\ref{Corr}) is justified by the fact that for a general uniform gradient (\ref{generalGradient}) the following expression 
\begin{equation}
\langle x B_x + y B_y \rangle = -\frac{1}{2} \frac{\partial B_z}{\partial z} \langle x^2+y^2 \rangle = -\frac{R^2}{4} \frac{\partial B_z}{\partial z}
\end{equation}
holds in a cylindrical trap of radius $R$. 
In the general case, to account for localized magnetic field, we define the enhancement factor $1+E$ as
\begin{equation}
\langle x B_x + y B_y \rangle = -\frac{R^2}{4} \langle \frac{\partial B_z}{\partial z} \rangle \left( 1+ E \right).
\label{EnFact}
\end{equation}
This enhancement factor strongly depends on the proximity of the small scale inhomogeneity.
Let us consider the case of a field induced by a dipole of magnetic moment $m = m e_z$ situated below the bottom electrode, on the axis of the cylindrical trap. 
The field reads, in cylindrical coordinates centered at the dipole position: 
\begin{eqnarray}
B_z    & = & \frac{p}{d^3} \left( \frac{3z^2}{d^2} - 1 \right) \\
B_\rho & = & \frac{p}{d^3} \ \frac{3 z \rho}{d^2}
\end{eqnarray}
where $p = \mu_0 m/4\pi$ is the strength of the dipole, 
$d = \sqrt{z^2 + \rho^2}$ is the distance from the dipole.
One can calculate the gradient
\begin{equation}
\frac{\partial B_z}{\partial z} = p \left( \frac{9z}{d^5} - \frac{15 z^3}{d^7} \right)
\end{equation}
and its volume average
\begin{equation}
\langle \frac{\partial B_z}{\partial z} \rangle = \frac{2 p}{H} \left( ((z_d+H)^2 + R^2)^{-3/2} - (z_d^2 + R^2)^{-3/2} \right)
\end{equation}
where $z_d$ is the distance of the dipole from the bottom electrode. 
Now the integral of the left hand side of eq. (\ref{EnFact}) is
\begin{equation}
\langle xB_x + yB_y \rangle = \langle \rho B_\rho \rangle = \frac{1}{\pi R^2 H} \int_{z_d}^{z_d+H} dz \int_0^R 2 \pi \rho d\rho \ \rho B_\rho
\end{equation}
evaluating the integral gives: 
\begin{eqnarray}
\langle xB_x + yB_y \rangle = \frac{2p}{R^2 H}  && \left( 2 H + \frac{R^2 
 + 2 z_d^2}{\sqrt{R^2 + z_d^2}} \right. \\ \nonumber && ~ \left. - \frac{R^2 + 2 (z_d+H)^2}{\sqrt{R^2 + (z_d+H)^2}} \right).
\end{eqnarray}
Then one can calculate analytically the enhancement factor: 
\begin{equation}
\label{AnalyticDipole}
1+E = - \frac{4}{R^2} \frac{\langle \rho B_\rho \rangle}{\langle \frac{\partial B_z}{\partial z} \rangle}. 
\end{equation}

As shown in fig. \ref{compHarrisPendlebury}, the analytical result derived here reproduces very well earlier numerical studies \cite{HarrisPendlebury}. 

\begin{figure}
\begin{center}
\includegraphics[width=0.98\linewidth]{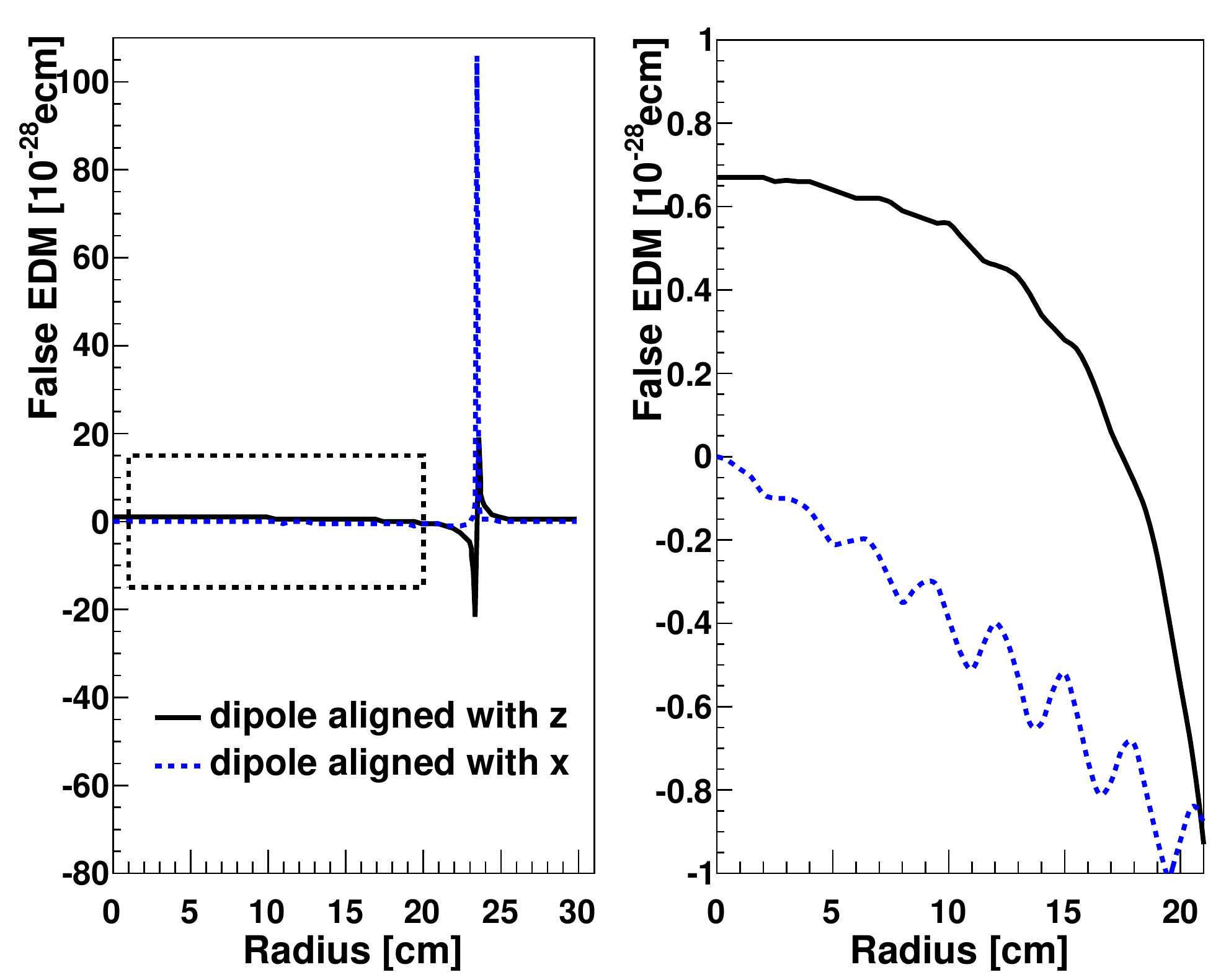}
\caption{
False neutron EDM signal due to a dipole of strength $p = 0.5 \ {\rm nT cm}^3$ at the lower surface of the cell for different radial positions and different orientations. 
The left hand side plot shows the main contribution around the circumference of the cell. 
The right hand side plot is a zoom for smaller radii.  
\label{FalseEDMdipole}
}
\end{center}
\end{figure}

The improved theory can now predict the enhancement factor even when the dipole is at the surface of the cell, which corresponds to the worst case. 
We have calculated the systematic effect on the neutron EDM due to the presence of a dipole at the surface that remains after the correction (\ref{Corr}). 
When the dipole is not situated at the center of the electrode, we did not find analytical expressions for the volume average (\ref{dFalseHg}), the volume integrals have been calculated with numerical methods. 
The result is shown in fig. \ref{FalseEDMdipole}, as a function of the position of the dipole on the bottom electrode and for two orientations of the dipole. 
We found that a dipole situated at the circumference is unfortunately particularly effective at generating a false EDM. 
This result can be used to set the requirements on the control of magnetic impurities in an experiment searching for the neutron EDM. 

\section{Conclusion}

In the context of the search for the neutron EDM with an atomic comagnetometer, most of the systematic effects are related with the geometric phase shift. 
In the nonadiabatic limit which describes a comagnetometer at low pressure, 
an improved theory valid for any geometry of the trap and arbitrary shape of the magnetic field was developed. 
It permits to calculate the false EDM generated by the geometric phase shift of mercury using the general formula (\ref{dFalseHg}) 
thus avoiding recoursing to heavy spin-tracking Monte-Carlo codes.

\section*{Acknowledgements}

We are grateful to the PSI nEDM collaboration which inspired this work,
especially to K.~Kirch and D.~Rebreyend. 

\appendix

\section{Spin relaxation and frequency shifts: an elementary derivation}
\label{derivation}

When an ensemble of spin-polarized particles evolve in a weakly fluctuating magnetic field, 
two important phenomena occur: spin-relaxation and shift of the resonance line. 
The situation is particularly relevant for a gas of polarized particles evolving in a static, but inhomogeneous magnetic field. 
In this case each particle sees effectively a time dependent magnetic field, which differ from particle to particle. 
Here, an elementary derivation of the basic formulas is given for the spin 1/2 case. 
The classic derivations are usually done within the density matrix formalism, as a special case of the general theory of relaxation developed by Redfield and others (see for example \cite{Abragam} for such a derivation). 
Here we derive the same results using the classical equations of motion of single magnetic moments ${\bf M}$ with gyromagnetic ratio $\gamma$ evolving in a weakly fluctuating magnetic field {\bf B}: 
\begin{equation}
\frac{d {\bf M}}{dt} = - \gamma \, {\bf B} \times {\bf M}.
\end{equation}
We assume that the magnetic field can be written in the following form: 
\begin{equation}
{\bf B}(t) = B_0 {\bf e}_z + \frac{1}{\gamma} {\boldsymbol \omega}(t) \quad {\rm with} \quad \langle {\boldsymbol \omega}(t) \rangle = {\bf 0}.
\end{equation}
It is indeed always possible to define the $z$ axis to coincide with the direction of the average field. 
Here $\langle X \rangle$ denotes the ensemble average of the quantity $X$. 
We proceed and define the magnetic moments in the frame rotating around the $z$ axis at the frequency $\omega_0 = \gamma B_0$: 
\begin{equation}
{\bf m} = R(t) \, {\bf M}, 
\end{equation}
with $R(t)$ being the rotation matrix:
\begin{equation}
R(t) = \left( \begin{array}{ccc}
\cos(\omega_0 t) & - \sin(\omega_0 t) & 0 \\
\sin(\omega_0 t) & \cos(\omega_0 t) & 0 \\
0 & 0 & 1 \end{array} \right).
\end{equation}
Then the magnetic moment of a given particle seen in the rotating frame evolves as:
\begin{equation}
\label{Bloch}
\frac{d {\bf m}}{dt} = - \gamma \, {\bf b} \times {\bf m}
\end{equation}
where ${\bf b}(t) = R(t) \, {\boldsymbol \omega}(t) / \gamma$ is the fluctuating magnetic field in the rotating frame. 
We are interested in deriving a simple evolution equation of the average magnetization $\langle {\bf m} \rangle$. 
We start with the ensemble average of eq. (\ref{Bloch})
\begin{equation}
\frac{d \langle {\bf m} \rangle}{dt} = - \gamma \, \langle {\bf b} \times {\bf m} \rangle
\end{equation}
which can be successively integrated: 
\begin{eqnarray}
\frac{d \langle {\bf m} \rangle}{dt} = & - & \gamma \, \langle {\bf b}(t) \times {\bf m}(0) \rangle \\
\nonumber
& + & \gamma^2 \int_0^t dt' \langle {\bf b}(t) \times \left( {\bf b}(t') \times {\bf m}(t') \right) \rangle.
\end{eqnarray}
The first term vanishes because we assume the absence of correlation between the initial direction of ${\bf m}$ and the future magnetic field ${\bf b}$ for individual particles. 
The variations of $\langle {\bf m} \rangle$ will be considered as a perturbation, which is valid in the limit of small fluctuating fields $\langle {\bf b}^2 \rangle \ll B_0^2$. 
The perturbative approach consists in substituting in the second term ${\bf m}(0)$ in place of ${\bf m}(t')$. 
Then we arrive at the first order equation, by introducing a new variable $\tau = t - t'$: 
\begin{eqnarray}
\label{relaxationEquation}
\frac{d \langle {\bf m} \rangle}{dt} & = & 
\gamma^2 \int_0^t d \tau \langle {\bf b}(t) \times \left( {\bf b}(t-\tau) \times {\bf m} \right) \rangle \\
\nonumber
& \equiv & - \, \Gamma \,\langle {\bf m} \rangle
\end{eqnarray}
where we define the \textit{relaxation matrix} $\Gamma$. 
Using the triple product expansion, we obtain an expression of the relaxation matrix involving autocorrelations of the components of the fluctuating field: 
\begin{eqnarray}
\label{Gammaij}
\Gamma_{ij} = & &  \gamma^2 \int_0^t d \tau \langle {\bf b}(t) {\bf b}(t-\tau) \rangle \\
\nonumber
& - & \gamma^2 \int_0^t d \tau \langle b_i(t - \tau) {b_j}(t) \rangle.
\end{eqnarray}
This matrix should be decomposed in the following form, with transparent meaning: 
\begin{equation}
\label{GammaParam}
\Gamma = \left( \begin{array}{ccc}
\Gamma_2       & - \delta \omega  & 0 \\
\delta \omega  & \Gamma_2         & 0 \\
0              & 0                & \Gamma_1 \end{array} \right) 
+ \left( \begin{array}{ccc}
\gamma   & \delta   & \beta_y \\
\delta   & -\gamma  & \beta_x \\
\alpha_y & \alpha_x & 0 \end{array} \right).
\end{equation}
The second matrix in the right hand side does not lead to first order effects because it corresponds to oscillating terms, at the frequency $2 \omega_0$, when transforming the equation (\ref{relaxationEquation}) back to the laboratory frame. 
All important phenomenological consequences of the fluctuating field are encoded in the first matrix, where $\Gamma_1 = \frac{1}{T_1}$ 
is the longitudinal spin-relaxation rate, $\Gamma_2 = \frac{1}{T_2}$ is the transverse spin-relaxation rate, and $\delta \omega$ represents the shift of the resonant frequency. 

Now, one can derive from (\ref{Gammaij}) and (\ref{GammaParam}) the relevant terms of the relaxation matrix, 
in terms of the autocorrelation of the fluctuating field expressed in the laboratory frame. 

\small
\begin{eqnarray}
\label{T1}
\Gamma_1 =   \int_0^\infty d\tau \cos(\omega_0 \tau) \langle \omega_x(0) \omega_x(\tau) + \omega_y(0) \omega_y(\tau) \rangle \\ 
\nonumber
	 +  \int_0^\infty d\tau \sin(\omega_0 \tau) \langle \omega_y(0) \omega_x(\tau) - \omega_x(0) \omega_y(\tau) \rangle 
\end{eqnarray}

\begin{equation}
\label{T2}
\Gamma_2 = \frac{1}{2} \, \Gamma_1 + \int_0^\infty d\tau \langle \omega_z(0) \omega_z(\tau) \rangle
\end{equation}

\begin{eqnarray}
\label{shift}
\delta \omega = \frac{1}{2}  \int_0^\infty d \tau \cos(\omega_0 \tau) \langle \omega_x(0) \omega_y(\tau) - \omega_y(0) \omega_x(\tau) \rangle \\
\nonumber
	+	\frac{1}{2} \int_0^\infty d \tau \sin(\omega_0 \tau) \langle \omega_x(0) \omega_x(\tau) + \omega_y(0) \omega_y(\tau) \rangle.
\end{eqnarray}
\normalsize

To get the final expressions (\ref{T1}) (\ref{T2}) (\ref{shift}), further reasonable assumptions have been made. 
It has been assumed that the system is stationary in the statistical sense, that is, the correlations depends only on the time difference: 
\begin{equation}
\langle b_i(t) b_j(t-\tau) \rangle = \langle b_i(\tau) b_j(0) \rangle. 
\end{equation}
Also, it has been assumed that the integrals in eq. (\ref{Gammaij}) can be extended to infinite time, 
which is justified because the correlations decay with a time much shorter than the observation time. 
When disregarding the transient regime immediately after the time zero for an observation, the relaxation matrix is time-independent. 
As a final remark, a similar calculation has been done in the density matrix formalism to obtain the frequency shift in \cite{LamoreauxGolub}, 
which agrees with (\ref{shift}) after correcting \cite{LamoreauxGolub} for a sign error.

\end{document}